\documentclass[letter]{aa} 
\usepackage{graphicx}	
\usepackage{txfonts}
\begin{document} 

    \title{Are ultra-diffuse galaxies Milky Way-sized?}
   \author{Nushkia Chamba,
          \inst{1, 2}\thanks{email: chamba@iac.es}
          Ignacio Trujillo
          \inst{1, 2}
          \and
          Johan H. Knapen
          \inst{1, 2}
          }
   \institute{Instituto de Astrof\'isica de Canarias, c/ V\'ia L\'actea s/n, E-38205, La Laguna, Tenerife, Spain
         \and
             Departmento de Astrof\'isica, Universidad de La Laguna, E-38200, La Laguna, Tenerife, Spain
             }

   \date{Received YYY; Accepted XXX.}

\titlerunning{Are ultra-diffuse galaxies Milky Way-sized?}
\authorrunning{Nushkia Chamba, Ignacio Trujillo \& Johan H. Knapen}

\abstract
{Now almost 70 years since its introduction, the effective or half-light radius has become a very popular choice for characterising galaxy size. However, the effective radius measures the concentration of light within galaxies and thus does not capture the intuitive definition of size which is related to the edge or boundary of objects. For this reason, we aim to demonstrate the undesirable consequence of using the effective radius to draw conclusions about the nature of faint `ultra-diffuse galaxies' (UDGs) when compared to dwarfs and Milky Way-like galaxies. Instead of the effective radius, we use a measure of galaxy size based on the location of the gas density threshold required for star formation. Compared to the effective radius, this physically motivated definition places the sizes much closer to the boundary of a galaxy. Therefore, considering the sizes and stellar mass density profiles of UDGs and regular dwarfs, we find that the UDGs have sizes that are within the size range of dwarfs. We also show that currently known UDGs do not have sizes comparable to Milky Way-like objects. We find that, on average, UDGs are ten times smaller in extension than Milky Way-like galaxies. These results show that the use of size estimators sensitive to the concentration of light can lead to misleading results.}


\maketitle
\section{Introduction}
Faint galaxies with large effective radii have been known since the 1980s \citep[e.g.][]{1984bruno, 1988impey, 1991bothun, 1997dalcanton}, but more recently, the name  ultra-diffuse galaxies \citep[UDGs;][]{2015giantgalaxies} has been coined for galaxies with very similar characteristics. These are galaxies of low stellar density, defined to have low central surface brightness ($\mu_g(0)>24$\,mag/arcsec$^2$) and an effective radius ($R_{\rm e}$) of over 1.5\,kpc \citep[{$R_{\rm e}$} is the radius which encloses half the total flux from a galaxy;][]{1948devaucouleurs}. The question of whether UDGs represent a separate class of galaxies is still under debate. Currently, known UDGs that have been discovered in clusters \citep{2015koda,2015mihos, 2015munoz, 2016vanderburg, 2017roman, 2017venhola, 2018pina}, in groups \citep{2017s82udgs, 2018cohen}, and in the field \citep{2017bellazzini, 2019prole} can have $R_{\rm e}$ as large as 5 kpc which is comparable to that of large (i.e. giant) Milky Way (MW)-like galaxies. This fact has been used to suggest that UDGs are `failed' giants \citep{2015giantgalaxies}. As $R_{\rm e}$ captures (at most) the central parts of giant galaxies, whether this radius can be used to fairly compare the sizes of UDGs to the more massive galaxies is questionable. \par

The reason why $R_{\rm e}$ is incapable of reaching the boundaries of massive galaxies is that according to its definition it depends on how the light is concentrated in these objects. Therefore, if one considers that the sizes of galaxies are indicated by the location of their edges or boundaries (similar to everyday objects), then $R_{\rm e}$ is undeniably a poor measurement of size. However, the idea of associating the sizes of galaxies to the location of their boundaries (or something very close to them) is not recent. Galaxy size has also been measured using limiting surface brightness isophotes, such as for example $R_{25}$ \citep{1936redman} or the Holmberg radius \citep[{$R_{\rm H}$};][]{1958holmberg}, to characterise the maximum area that galaxies spanned on the photographic plates of that era. However, similar to the effective radius, the isophotal radii were also initially defined for operational purposes and do not directly encompass any physical meaning. In spite of this, isophotal radii \citep[and their variants; see e.g.][]{2012hall} as well as sizes based on light concentration, for example $R_{\rm e}$, $R_{90}$ \citep{2011nair}, and $R_{80}$ \citep{2019miller}, are being used in important scaling relations such as the fundamental plane \citep{1987DS}, size--stellar mass \citep{2003shen}, or the size--virial radius relation \citep{2013kravtsov}, to study the history and formation of galaxies. \par

With the aim of finding a physically motivated characterisation of galaxy size, Trujillo, Chamba \& Knapen (submitted, hereafter TCK19) proposed a size parameter based on the location of the gas density threshold for star formation in galaxies. We found that the size--stellar mass relation with this measure for size has an intrinsic dispersion of only $\sim$0.06 dex, which is three times smaller than that of the relation with $R_{\rm e}$ as galaxy size ($\sim$0.18 dex), over five orders of magnitude in stellar mass $10^{7}\,M_{\odot}<M_{\star}<10^{12}\,M_{\odot}$. The proposed parameter is also able to capture the boundaries of the stellar distribution of galaxies and can thus represent how large or small these objects are, in contrast to the effective radius. We refer the reader to TCK19 for an in-depth discussion on the fundamental meaning of these results and why such a size definition is different from the ones that only measure the extent of galaxies down to a given surface brightness level (e.g. $R_{\rm H}$ or $R_{25}$). \par

To complement the results presented in TCK19, in this \textit{Letter} we study the implications of using the effective radius as a size measure for UDGs; and how this affects our understanding of these galaxies. We compute the physically motivated size parameter defined in TCK19 for a sample of UDGs and compare their sizes to those of dwarfs with stellar masses  10$^{7}\,M_{\odot}$$\leq$$M_{\star}$$\leq$10$^{8.5}\,M_{\odot}$ as well as to the sizes of MW-like galaxies (10$^{10}\,M_{\odot}$<$M_{\star}$<10$^{11}\,M_{\odot}$) studied in TCK19. Throughout this work we assume a standard $\Lambda$CDM cosmology with $\Omega_m$=0.3, $\Omega_\Lambda$=0.7 and $H_0$=70 km s$^{-1}$ Mpc$^{-1}$. \par

\section{Data and sample selection}

In order to have a homogeneous dataset of dwarfs and UDGs both in depth and filter coverage, we use publicly available background-rectified imaging data in the $g$ and $r$-bands of the deep IAC Stripe 82 Legacy Project\footnote{\protect\url{http://research.iac.es/proyecto/stripe82/}} \citep[hereafter IAC Stripe82,][]{2016s82legacy, 2018s82rectified}. The UDGs are taken from \citet{2017s82udgs} and \citet{2017ugc2162}. For completeness, two iconic UDGs outside the Stripe 82 footprint were also added to the sample. Imaging data for DF44 \citep{2015df44}, a representative example of a UDG with a large $R_{\rm e}$, was obtained from the Gemini archive (GN-2016A-FT-18,PI: P. van Dokkum) and DECaLS data\footnote{\protect\url{http://portal.nersc.gov/project/cosmo/data/legacysurvey/dr7/coadd/195/1952p270/}} was used for [KKS2000]04 (popularized as NGC1052-DF2)\footnote{At a distance of 13 Mpc \citep{2019trujillo}, [KKS2000]04 no longer satisfies the criterion ($R_{\rm e}$ > 1.5 kpc) to be defined as a UDG. Nevertheless, due to the popularity of this object after being reported as a `galaxy lacking dark matter' \citep{2018naturedok}, we include it in our sample for analysis.}. The control sample analysed consists of 155 dwarf galaxies with stellar masses in the range 10$^7\,M_{\odot}$$\leq$$M_{\star}$$\leq$10$^{8.5}\,M_{\odot}$ studied in TCK19. We focus on this stellar mass regime as it overlaps with the mass range of the selected UDGs for this work. Galaxies with stellar masses in the range of 10$^{10}\,M_{\odot}$<$M_{\star}$<10$^{11}\,M_{\odot}$ from the TCK19 catalogue (449 objects) are also selected to represent MW-like systems. \par 
The IAC Stripe82 and DECaLS images are of similar depth with a limit in surface brightness of $\mu_g$~=~29.1\,mag/arcsec$^2$\,(3$\sigma$;10$\times$10\,arcsec$^2$). The depth of the Gemini coaddition is $\mu_g\sim30$\,mag/arcsec$^2$\,(3$\sigma$;10$\times$10\,arcsec$^2$). Only the $g$- and $i$-band data were available for [KKS2000]04.\par 
Ultra-diffuse galaxies R2 and R3 from \citet{2017s82udgs} were removed from the UDG sample due to light contamination in the galaxy outskirts produced by surrounding bright sources and/or stars. Therefore, our final sample includes 12 UDGs. None of the dwarf galaxies in our control sample satisfy the criteria for a UDG as in all the cases $\mu_g(0)<24$\,mag/arcsec$^2$.

\section{Method} \label{section:Method}

The entire analysis of this study was carried out on individual image stamps with dimensions of 100$\times$100\,kpc$^2$ (for the dwarfs and UDGs) and 600$\times$600\,kpc$^2$ (for the massive galaxies) in the rest frame of each galaxy. The scattered light from point sources was removed from the IAC Stripe82 image stamps using our extended ($\sim$ 8\,arcmin radius) point spread functions for this telescope\footnote{\protect\url{http://research.iac.es/proyecto/stripe82/pages/advanced-data-products/the-sdss-extended-psfs.php}} \citep{2019raul}. All sources surrounding the galaxy of interest in the image were then masked using \texttt{MTObjects} \citep{2016mto}, setting \texttt{move\_up = 0.3}. \par

To derive the surface brightness profiles of the galaxies in our sample, the axis-ratio ($q$) and position angle (PA) of the galaxies were obtained by fitting an ellipse to an average isophote of 26 mag/arcsec$^2$ in the $g$-band images. The centre, $q,$ and PA of each galaxy were then visually verified prior to further analysis. These parameters were fixed and elliptical annuli were used to create the radial profile of each galaxy as well as its growth curve in flux which is needed to determine $R_{\rm e}$. \par
The $g-r$ colour ($g-i$ for DF44) profiles were derived from the surface brightness profiles and converted to mass-to-light ratio ($M/L$) profiles in $g$ using the relationships from \citet{2015roediger}. These $M/L$ and surface brightness profiles in the $g$-band for all galaxies were then converted to stellar mass density ($\Sigma_{\star}$) profiles \citep[see Eq. 1 in][]{2008bakos} and used to ascertain our size parameter (TCK19). The profiles were also integrated up to the $\mu_g=29$\,mag/arcsec$^2$ isophote to derive the stellar masses of the galaxies, $M_{\star}$. Various stellar density thresholds (within the limit in depth of the images used) can easily be determined from such profiles. Here, we use $R_1$, the radius at which $\Sigma_{\star}=1\,M_{\odot}/$pc$^2$, as a proxy for the location of the gas density threshold for star formation (see TCK19). For details on the background subtraction of the data, correction of the profiles due to the inclination effect and Galactic extinction, and an estimation of the uncertainties related to our measurements (stellar mass and background) we refer the reader to TCK19. \par 
All of our measurements for the UDGs are provided in Appendix A. The measurements for the dwarf sample can be found in the online version of TCK19. For comparison, we also show the distributions of the UDGs and dwarfs using an isophotal size indicator, the Holmberg radius ($R_{\rm H}$), in Appendix \ref{sec:holmbergradius}. Lacking the B-band, we used the isophote at 26\,mag/arcsec$^2$ in the $g$-band as a proxy for $R_{\rm H}$.

\begin{figure*}
    \includegraphics[width=1.0\textwidth]{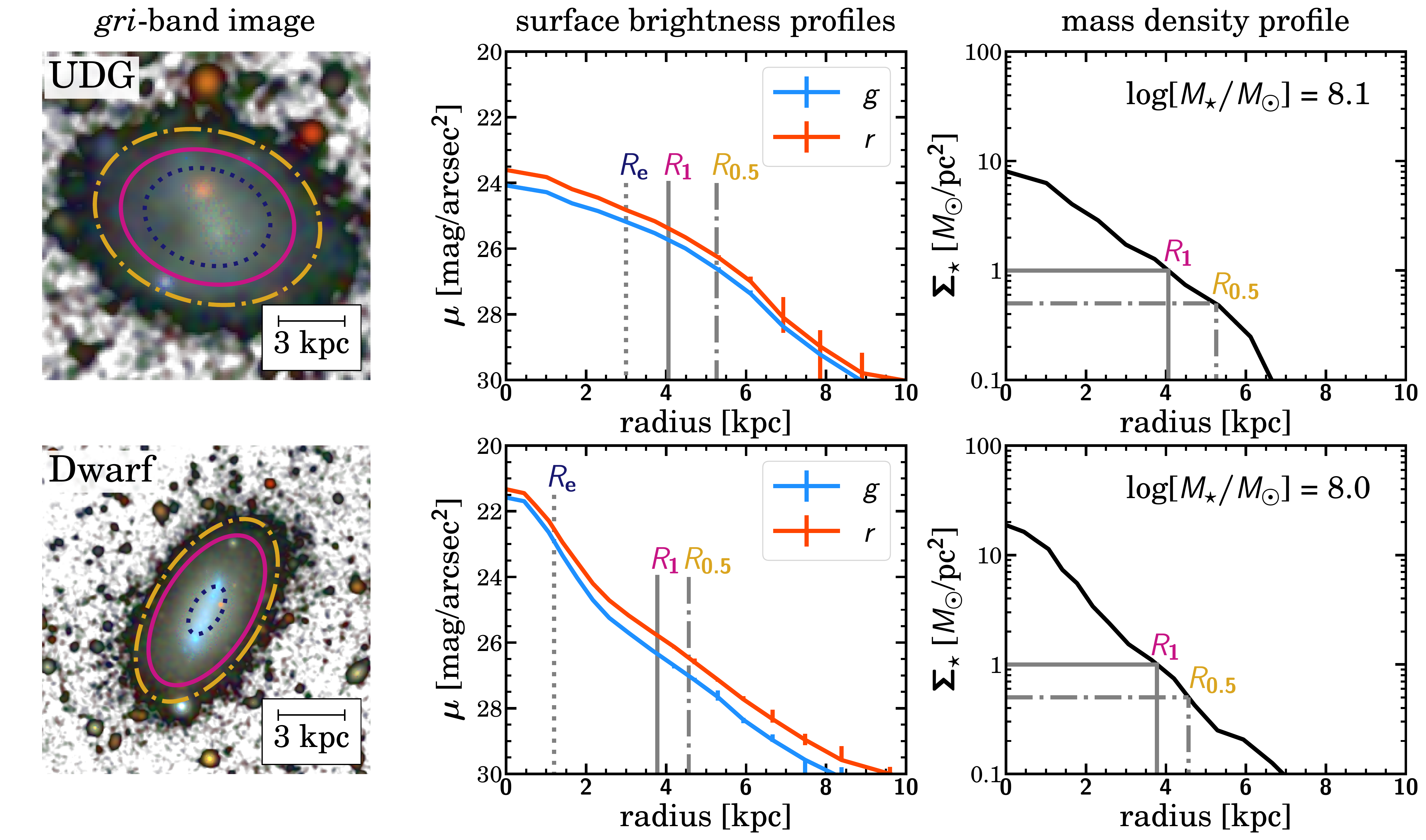}
    \caption{Illustration of the consequence of using $R_{\rm e}$ as galaxy size for UDGs and dwarf galaxies. Here we show two galaxies of similar stellar mass at the same physical scale: UDG-B5 (top) and a representative dwarf galaxy (SDSS J224114.12-003715.0, bottom). The colour image is the \textit{gri}-band composite with a grey-scaled background for contrast and contours showing $R_{\rm e}$ (dotted), $R_1$, (solid) and $R_{0.5}$ (dot-dashed). The surface brightness and stellar mass density profiles derived for both galaxies are also shown.}
    \label{fig:egprofiles}
\end{figure*}

\begin{figure*}
    \centering 
    \includegraphics[width=1.0\textwidth]{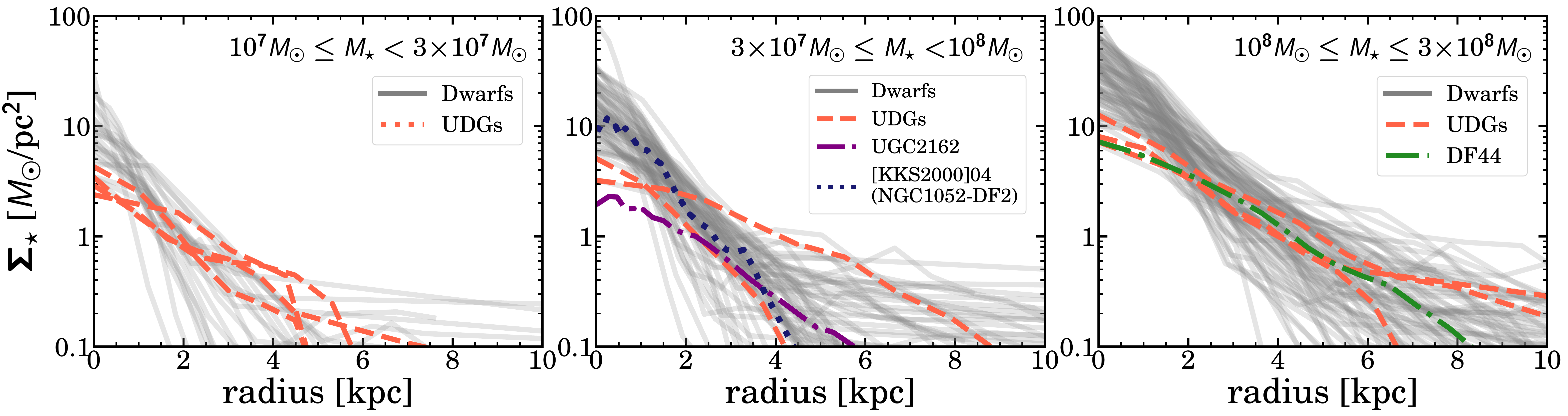}
    \caption{Stellar mass density profiles of dwarfs (grey) and UDGs (colours) belonging to three stellar mass bins (left to right).}
    \label{fig:allprofiles}
\end{figure*}

\begin{figure*}[h]
    \hspace{0.05cm}
    \includegraphics[width=1.0\textwidth]{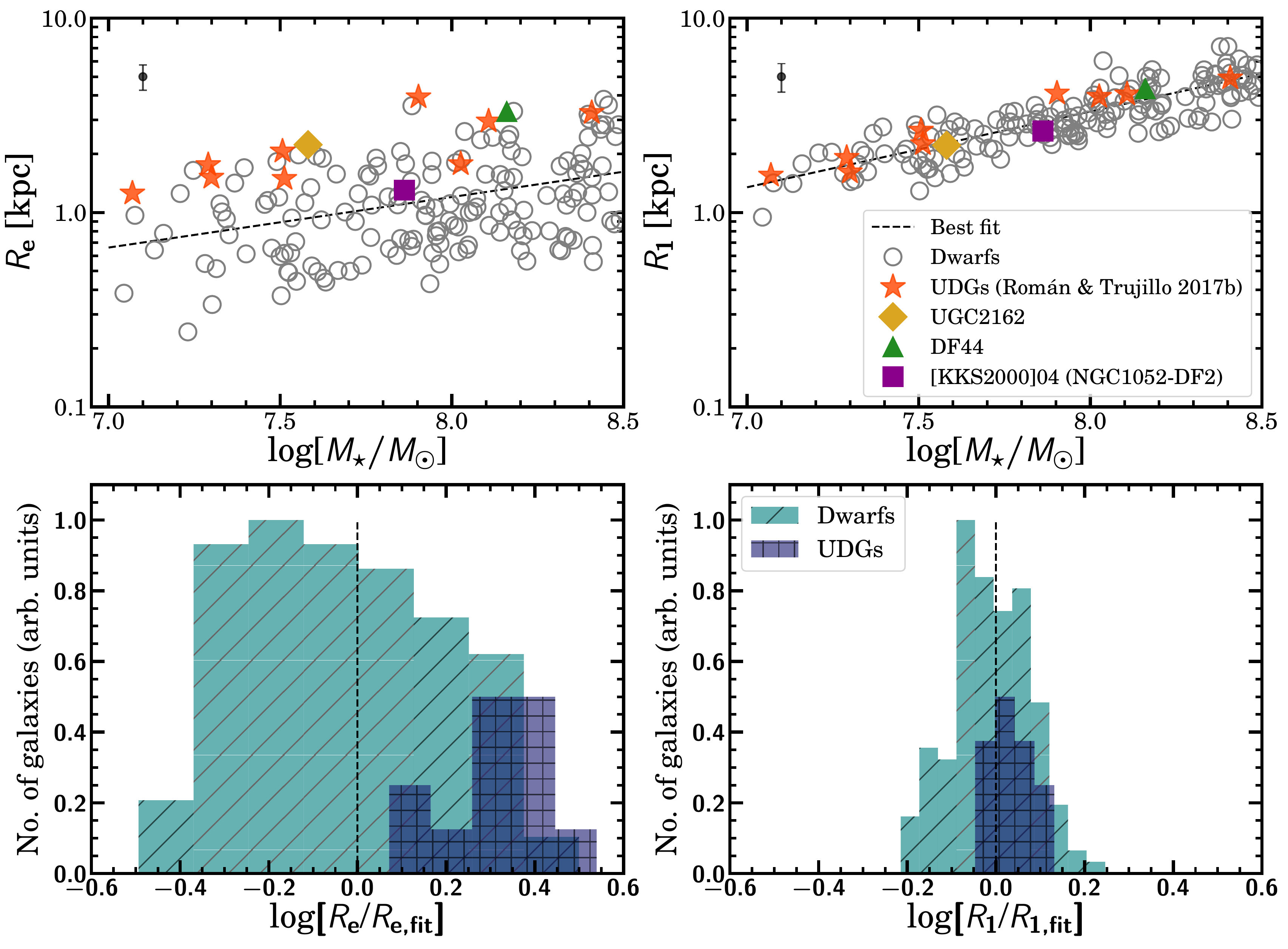}
    \caption{Comparison between $R_{\rm e}$ and the physically motivated size parameter for UDGs and dwarfs. \textit{Top}: $R_{\rm e}$--stellar mass relation (left) and the $R_1$--stellar mass relation (right) for dwarfs (grey) and UDGs (colours). The best-fit line of each relation for the dwarf sample is also over-plotted. The upper left corner of each plot shows the typical uncertainty in our measurements (see TCK19). \textit{Bottom}: Histograms showing the distribution of $R_{\rm e}/R_{\rm e, fit}$ (left) and $R_1/R_{1,\rm fit}$ (right) where `fit' refers to the best-fit line of each relation for the dwarf sample.}
    \label{fig:relation}
\end{figure*}

\begin{figure*}[h]
    \centering
        \includegraphics[width=1.0\textwidth]{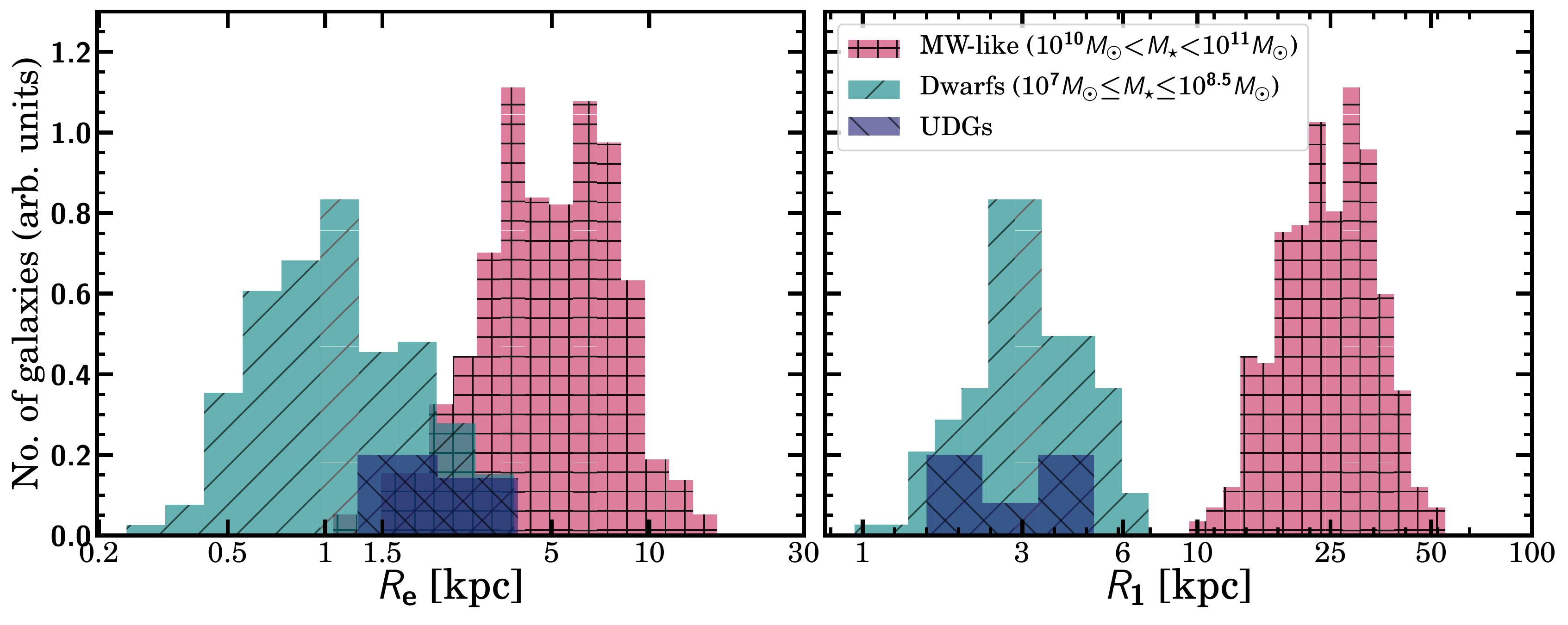}
    \caption{Histograms showing the size distribution of UDGs, dwarfs, and MW-like galaxies. In $R_{\rm e}$ (left), UDGs overlap with the dwarfs and MW-like systems in our sample and in $R_1$ (right), the UDGs clearly separate from the MW-like galaxies and overlap with the dwarfs. These results show that UDGs have the extensions of dwarfs.}
    \label{fig:histograms}
\end{figure*}

\begin{figure*}
    \centering
    \includegraphics[width=0.9\textwidth]{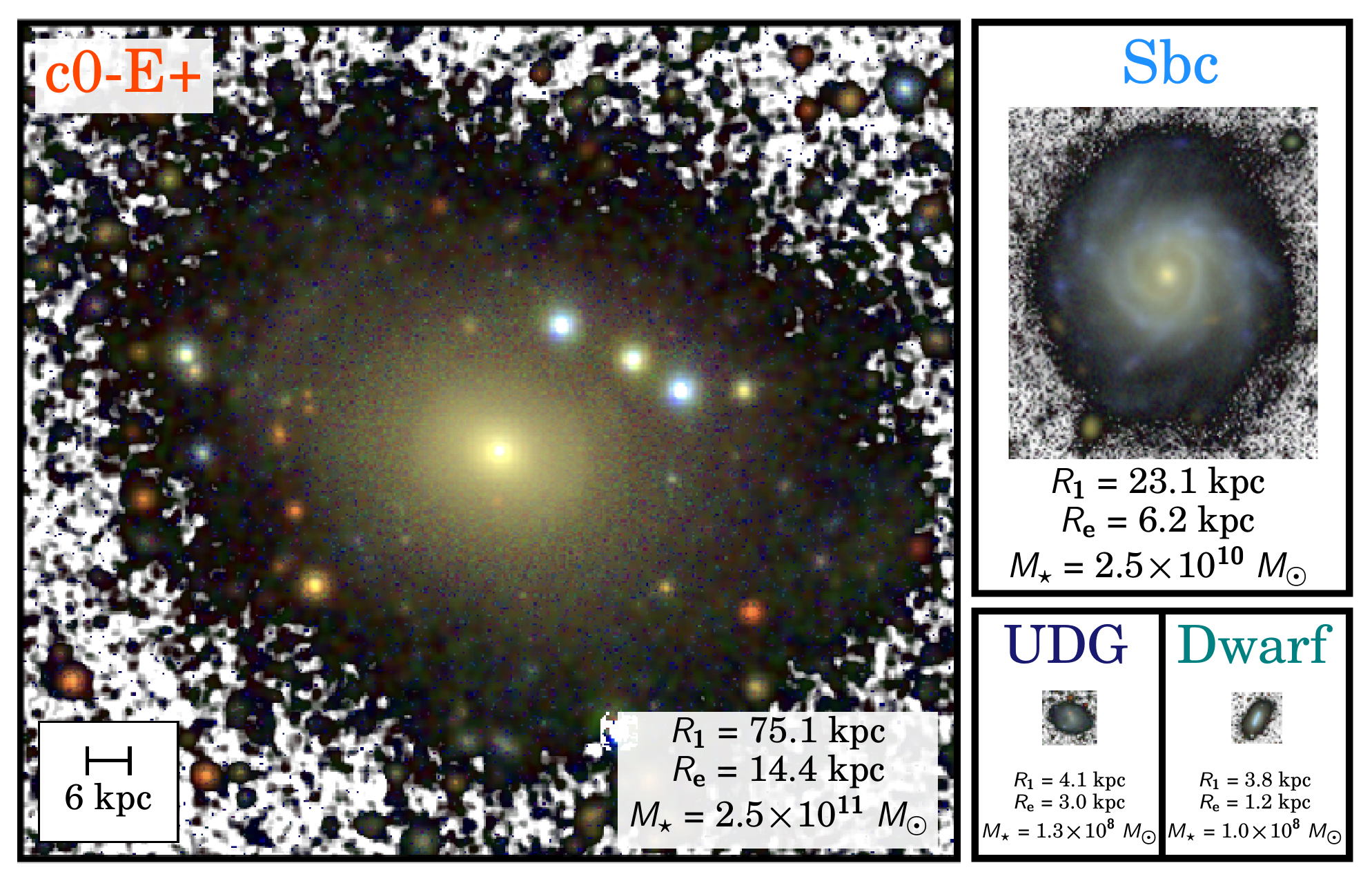}
    \vspace{0.5mm}
    \includegraphics[width=0.7\textwidth]{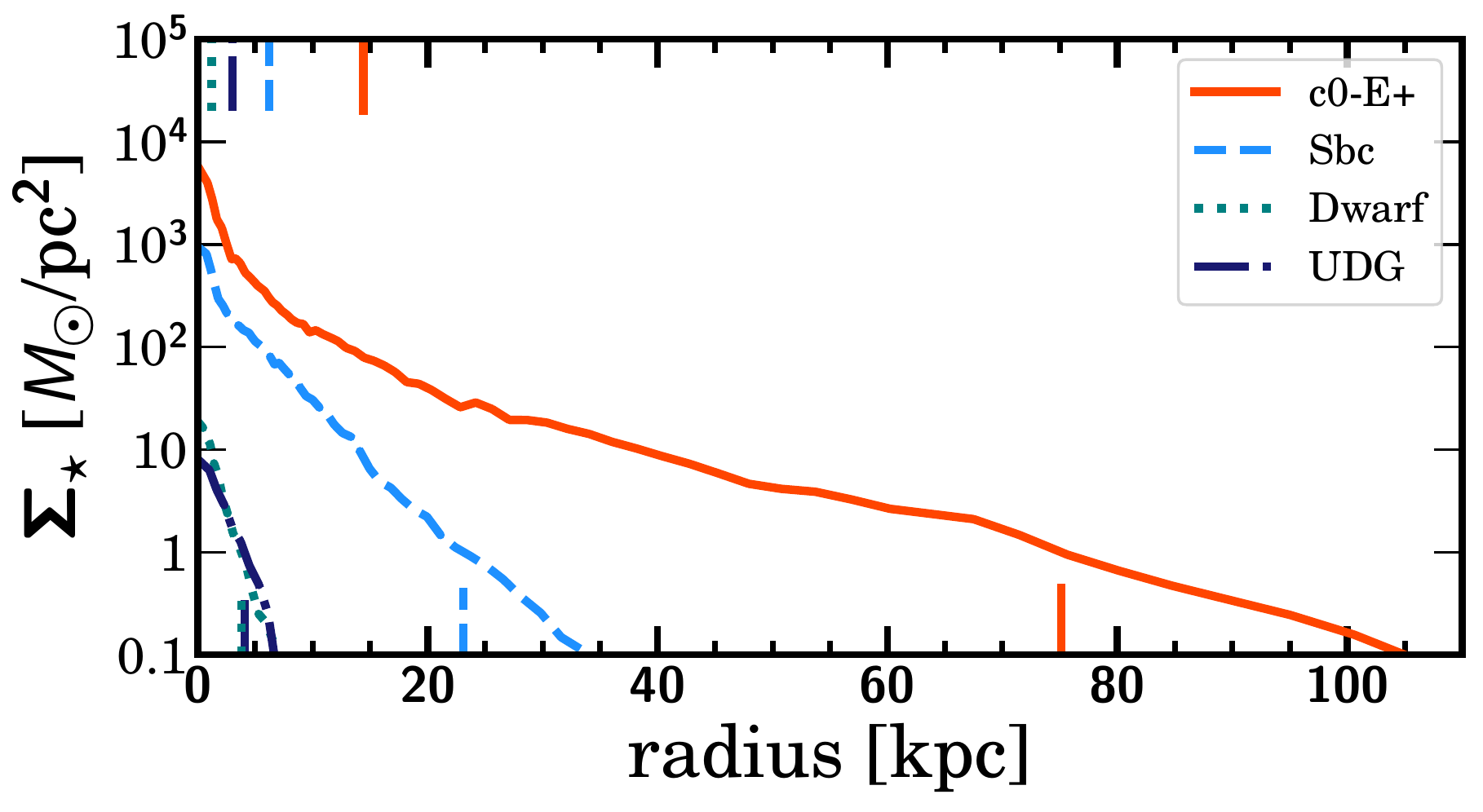}
    \caption{\textit{Top:} Representative galaxies of different stellar masses: a giant elliptical ($M_{\star} \sim 2.5\times10^{11}\,M_{\odot}$), a Milky Way-like galaxy ($M_{\star} \sim 2.5\times10^{10}\,M_{\odot}$), a UDG and dwarf galaxy ($M_{\star} \sim 10^{8}\,M_{\odot}$). All galaxies are shown to the same physical scale and a similar depth in surface brightness. \textit{Bottom:} The stellar mass density profiles of the same galaxies. The coloured ticks in the upper $x$-axis mark the location of $R_{\rm e}$. Similar ticks in the lower $x$-axis mark the location of $R_1$.}
\label{fig:samedist}
\end{figure*}

\section{Results}

Figure \ref{fig:egprofiles} shows an example of a UDG and a representative dwarf galaxy (i.e. one that lies very close to the centre and best-fit line in both the observed $R_{\rm e}$-- and $R_1$--stellar mass relations). Both galaxies have similar stellar mass ($\sim$ $10^{8}$\,$M_{\odot}$). Their corresponding surface brightness and mass density profiles are shown, and the locations of $R_{\rm e}$ (dotted), $R_1$ (solid) and the radius where $\Sigma_{\star}=0.5$\,$M_{\odot}$/pc$^2$ (called $R_{0.5}$; dot-dashed, see Appendix \ref{sec:othersize}) are marked in the image and profiles. The reason why the effective radius of the UDG is large in comparison to that of the regular dwarf galaxy shown is because the dwarf galaxy has active star forming clumps in its central region. The presence of such clumps in these galaxies means that flux will be more concentrated in the centre which decreases the effective radius and increases their central surface brightness. Consequently, such dwarfs will not be characterised as a UDG. Similar clumps or bright regions are not usually present at the centre of UDGs which makes their effective radii larger compared to the majority of dwarfs. \par
This last point is further demonstrated in Fig. \ref{fig:allprofiles} where the $\Sigma_{\star}$ profiles of dwarf galaxies and UDGs are over-plotted in three panels, corresponding to galaxies in three stellar mass bins: 10$^7\,M_{\odot}\leq\,M_{\star}<3$$\times$$10^{7}\,M_{\odot}$, 3$\times$10$^7\,M_{\odot}\leq\,M_{\star}<10^{8}\,M_{\odot}$, and $10^{8}\,M_{\odot}\leq\,M_{\star}\leq\,3\times10^{8}\,M_{\odot}$. Ultra-diffuse galaxies tend to be less concentrated than regular dwarfs, with lower central densities by a factor of two to three. However, the global extensions of both types of galaxy are very much alike.

Figure \ref{fig:relation} shows the $R_{\rm e}$--stellar mass and $R_1$--stellar mass scaling relationships (top panels) and their distributions with respect to the best-fit lines (bottom panels). Three main features are: 1) The dispersion of the observed $R_1$--stellar mass plane (0.086\,$\pm$\,0.007 dex) is a factor of 2.5 smaller than that of the observed $R_{\rm e}$--stellar mass plane (0.213\,$\pm$\,0.014 dex) for dwarf galaxies (see also TCK19); 2) UDGs populate the upper portion of the $R_{\rm e}$-stellar mass plane. A simple Kolmogrov-Smirnov (KS) test using $R_{\rm e}/R_{\rm e, fit}$ gives an extremely small p-value of $2.1 \times 10^{-5}$. After removing two galaxies that have $R_{\rm e} < 1.5\,$kpc (according to our measurements) in our initial UDG sample, namely [KKS2000]04 (NGC1052-DF2) and UDG-R1 \citep[from][]{2017s82udgs}, the p-value decreases to 9.3$\times 10^{-6}$. Both values indicate that the null hypothesis --- that dwarf galaxies and UDGs in this sample arise from the same distribution in size --- can be rejected; and 3) UDGs are populated among the dwarf galaxies in the $R_1$--stellar mass plane, showing no evidence that the distributions in $R_1/R_{\rm 1, fit}$ of these galaxies are significantly different (p-value = 0.07). After removing the two galaxies with $R_{\rm e} < 1.5\,$kpc, the p-value increases to 0.09. Therefore, the null hypothesis cannot be rejected. The UDGs shown in this work have extensions that correspond to those of dwarfs. We repeated this exercise using another popular size indicator, the Holmberg radius, and found similar results (see Appendix \ref{sec:holmbergradius}). \par 
On average, the location of $R_1$ in surface brightness for galaxies in this mass range is $\mu_g(R_1) \sim 27$ mag/arcsec$^2$, but can be as faint as 28.5 mag/arcsec$^2$. Lower stellar mass densities are even more faint (e.g. $R_{0.5}$, see Appendix \ref{sec:othersize}), reinforcing the importance of high-quality deep images to conduct this work. \par 

Finally, we highlight our main results in Figs. \ref{fig:histograms} and \ref{fig:samedist}. Figure \ref{fig:histograms} demonstrates how using a physically motivated size parameter that captures the global extension of galaxies reveals the radical difference between the sizes of UDGs and MW-like galaxies (right panel), in contrast to the effective radius (left panel). While using $R_{\rm e}$ indicates that UDGs have similar extensions to MW-like galaxies, $R_1$ shows that the MW-like systems are, on average, ten times larger than the classical dwarfs and UDGs. In fact, the null hypothesis is completely rejected when the $R_1$ distributions of UDGs and MW-like galaxies are compared (see also Appendix \ref{sec:othersize}). This result is further illustrated in Fig. \ref{fig:samedist} where an elliptical galaxy (SDSS J223954.96-005918.97) reminiscent of M87, a MW-like spiral galaxy (SDSS J012015.34-002009.00), and the dwarf galaxy and UDG of Fig. \ref{fig:egprofiles} are shown to the same physical scale. The TType labels for the elliptical and spiral galaxies were taken from \citet{2010preethi}. We see that $R_1$ better represents the edges of galaxies compared to $R_{\rm e}$ and prevents any misleading notion about the actual extension of galaxies. We emphasise the fact that the galaxies are shown to a similar depth in surface brightness. Therefore, the strikingly different sizes are not a result of the quality of the imaging data.

\section{Discussion}
The aim of this study is to investigate how the effective radius, as a size measure for UDGs, affects our understanding of these galaxies. We illustrate that the effective radii of UDGs will generally always be larger compared to that of dwarf galaxies due to the absence of luminous clumps (or substructure like bulges in the case of more massive galaxies) in their central regions. The fact that the large effective radii of UDGs can be compatible with those of MW-like galaxies has led to the interpretation that UDGs are `Milky Way-sized' \citep[see e.g.][]{2015giantgalaxies, 2015koda}, when perhaps the more accurate statement is rather that `UDGs are less concentrated in light than dwarfs and MW-like galaxies'. For this reason, we adopted the physically motivated size measure that we developed in TCK19. As this size parameter is also better than the effective radius at representing how large or small galaxies are, we used it to fairly compare the sizes of UDGs with those of dwarfs and MW-like galaxies. Contrary to previous accounts, we demonstrate that the sizes of MW-like galaxies and UDGs are actually radically different. As a matter of fact, the sizes of UDGs with our definition (as well as with the Holmberg radius) are compatible with those of dwarfs. \par

However, while the KS test shows no evidence that the size distribution of the UDG and dwarf galaxy populations are different, most of the UDGs lie in the upper half of the $R_1$--stellar mass relation (Fig. \ref{fig:relation}). This could be related to the incompleteness in our dwarf control sample arising from the spectroscopic target selection criteria of the Sloan Digital Sky Survey (SDSS), requiring that the $r$-band Petrosian half-light surface brightnesses of targets are at least $\mu_{50} \leq 24.5$ mag/arcsec$^2$ \citep{2002strauss}. The lack of faint low-mass galaxies in our dwarf sample can also be seen in the stellar mass  density profiles in Fig. \ref{fig:allprofiles} where there are almost no dwarf galaxies with compatible central densities to UDGs. Any such bias due to spectroscopic incompleteness in our dwarf sample will also equally affect the $R_{\rm e}$--stellar mass plane. It is therefore an acceptable exercise to compare these relationships quantitatively. Were the control sample not affected by this potential incompleteness, the similarity between the dwarfs and UDGs in the distribution of their $R_1$ as well as $R_e$ would be even greater. This should be the case when future deep spectroscopic studies \citep[e.g.][]{2018tomas, 2018ferremateu} target more dwarf galaxies with lower central densities and essentially fill the upper portions of both these relations. \par

The use of the effective radius as a galaxy size measurement has also led to a confusion as to whether UDGs are associated to the dark matter haloes of MW-like (high-luminosity) or dwarf galaxies. Several analyses using simulations \citep[e.g.][]{2017dicintio, 2018chanT} and observations \citep[e.g.][]{2016beasleymar,2016beasleytrujillo,2018amoriscoapr} already support the idea that UDGs reside in haloes comparable to those of dwarfs. Our results lend further support to this idea. As the size of a galaxy is believed to be proportional to the virial radius of its halo \citep{2013kravtsov}, the fact that our UDG sizes agree with those of dwarfs strongly suggest that both types of galaxy occupy the same dark matter haloes. \par

\section{Conclusions}

We used a proxy for the gas density threshold for star formation as a size indicator for UDGs and dwarf galaxies. We compared the size distribution of these galaxies in a physically important parameter space --- the size--stellar mass plane --- and show that there is no evidence that the size distributions of UDGs and dwarfs are different. The UDGs have sizes that are within the size range of dwarfs. The same result holds using the Holmberg radius as a size indicator. If low-mass, extremely diffuse Milky Way-sized galaxies exist, then in our definition of size, they need to have a radius of about 25\,kpc. Such galaxies have not been found in present-day imaging surveys. Our results reinforce the importance of using physically meaningful properties in order to fairly compare classes of galaxies and draw conclusions about their nature. 

\bigskip
\footnotesize{\textit{Acknowledgements.} We thank the anonymous referee for providing useful comments. We also thank Javier Rom{\'a}n and Ra{\'u}l Infante for providing the extended SDSS point spread functions of all filters used in this work. We acknowledge Rodrigo Carrasco for kindly allowing us to use his reduction of the galaxy DF44 from the Gemini archive. NC thanks Caroline Haigh for providing the latest version of \texttt{MTObjects}. She also thanks Claudio Dalla Vecchia, Jorge S\'anchez Almeida and Mike Beasley for interesting conversations. We acknowledge support from the State Research Agency (AEI) of the Spanish Ministry of Science, Innovation and Universities (MCIU) and the European Regional Development Fund (FEDER) under the grants with references AYA2016-76219-P and AYA2016-77237-C3-1-P, from the European Union's Horizon 2020 research and innovation programme under Marie Sk\l odowska-Curie grant agreement No 721463 to the SUNDIAL ITN network, and from the Fundaci\'on BBVA under its 2017 programme of assistance to scientific research groups, for the project `Using machine-learning techniques to drag galaxies from the noise in deep imaging'. The IAC projects P/300624 and P/300724 is financed by the Ministry of Science, Innovation and Universities, through the State Budget and by the Canary Islands Department of Economy, Knowledge and Employment, through the Regional Budget of the Autonomous Community.}

\bibliographystyle{aa}
\bibliography{bibliography.bib}

\begin{thebibliography}{48}
\expandafter\ifx\csname natexlab\endcsname\relax\def\natexlab#1{#1}\fi

\bibitem[{{Amorisco} {et~al.}(2018){Amorisco}, {Monachesi}, {Agnello}, \&
  {White}}]{2018amoriscoapr}
{Amorisco}, N.~C., {Monachesi}, A., {Agnello}, A., \& {White}, S.~D.~M. 2018,
  \mnras, 475, 4235

\bibitem[{{Bakos} {et~al.}(2008){Bakos}, {Trujillo}, \& {Pohlen}}]{2008bakos}
{Bakos}, J., {Trujillo}, I., \& {Pohlen}, M. 2008, \apjl, 683, L103

\bibitem[{{Beasley} {et~al.}(2016){Beasley}, {Romanowsky}, {Pota}, {Navarro},
  {Martinez Delgado}, {Neyer}, \& {Deich}}]{2016beasleymar}
{Beasley}, M.~A., {Romanowsky}, A.~J., {Pota}, V., {et~al.} 2016, \apjl, 819,
  L20

\bibitem[{{Beasley} \& {Trujillo}(2016)}]{2016beasleytrujillo}
{Beasley}, M.~A. \& {Trujillo}, I. 2016, \apj, 830, 23

\bibitem[{{Bellazzini} {et~al.}(2017){Bellazzini}, {Belokurov}, {Magrini},
  {Fraternali}, {Testa}, {Beccari}, {Marchetti}, \& {Carini}}]{2017bellazzini}
{Bellazzini}, M., {Belokurov}, V., {Magrini}, L., {et~al.} 2017, \mnras, 467,
  3751

\bibitem[{{Bothun} {et~al.}(1991){Bothun}, {Impey}, \& {Malin}}]{1991bothun}
{Bothun}, G.~D., {Impey}, C.~D., \& {Malin}, D.~F. 1991, \apj, 376, 404

\bibitem[{{Chan} {et~al.}(2018){Chan}, {Kere{\v s}}, {Wetzel}, {Hopkins},
  {Faucher-Gigu{\`e}re}, {El-Badry}, {Garrison-Kimmel}, \&
  {Boylan-Kolchin}}]{2018chanT}
{Chan}, T.~K., {Kere{\v s}}, D., {Wetzel}, A., {et~al.} 2018, \mnras, 478, 906

\bibitem[{{Cohen} {et~al.}(2018){Cohen}, {van Dokkum}, {Danieli}, {Romanowsky},
  {Abraham}, {Merritt}, {Zhang}, {Mowla}, {Kruijssen}, {Conroy}, \&
  {Wasserman}}]{2018cohen}
{Cohen}, Y., {van Dokkum}, P., {Danieli}, S., {et~al.} 2018, \apj, 868, 96

\bibitem[{{Dalcanton} {et~al.}(1997){Dalcanton}, {Spergel}, {Gunn}, {Schmidt},
  \& {Schneider}}]{1997dalcanton}
{Dalcanton}, J.~J., {Spergel}, D.~N., {Gunn}, J.~E., {Schmidt}, M., \&
  {Schneider}, D.~P. 1997, \aj, 114, 635

\bibitem[{{de Vaucouleurs}(1948)}]{1948devaucouleurs}
{de Vaucouleurs}, G. 1948, Annales d'Astrophysique, 11, 247

\bibitem[{{Di Cintio} {et~al.}(2017){Di Cintio}, {Brook}, {Dutton},
  {Macci{\`o}}, {Obreja}, \& {Dekel}}]{2017dicintio}
{Di Cintio}, A., {Brook}, C.~B., {Dutton}, A.~A., {et~al.} 2017, \mnras, 466,
  L1

\bibitem[{{Djorgovski} \& {Davis}(1987)}]{1987DS}
{Djorgovski}, S. \& {Davis}, M. 1987, \apj, 313, 59

\bibitem[{{Emsellem} {et~al.}(2019){Emsellem}, {van der Burg}, {Fensch},
  {Je{\v{r}}{\'a}bkov{\'a}}, {Zanella}, {Agnello}, {Hilker}, {M{\"u}ller},
  {Rejkuba}, {Duc}, {Durrell}, {Habas}, {Lelli}, {Lim}, {Marleau}, {Peng}, \&
  {S{\'a}nchez-Janssen}}]{2019emsellem}
{Emsellem}, E., {van der Burg}, R. F.~J., {Fensch}, J., {et~al.} 2019, \aap,
  625, A76

\bibitem[{{Ferr{\'e}-Mateu} {et~al.}(2018){Ferr{\'e}-Mateu}, {Alabi}, {Forbes},
  {Romanowsky}, {Brodie}, {Pandya}, {Mart{\'\i}n-Navarro}, {Bellstedt},
  {Wasserman}, {Stone}, \& {Okabe}}]{2018ferremateu}
{Ferr{\'e}-Mateu}, A., {Alabi}, A., {Forbes}, D.~A., {et~al.} 2018, \mnras,
  479, 4891

\bibitem[{{Fliri} \& {Trujillo}(2016)}]{2016s82legacy}
{Fliri}, J. \& {Trujillo}, I. 2016, \mnras, 456, 1359

\bibitem[{{Hall} {et~al.}(2012){Hall}, {Courteau}, {Dutton}, {McDonald}, \&
  {Zhu}}]{2012hall}
{Hall}, M., {Courteau}, S., {Dutton}, A.~A., {McDonald}, M., \& {Zhu}, Y. 2012,
  \mnras, 425, 2741

\bibitem[{{Holmberg}(1958)}]{1958holmberg}
{Holmberg}, E. 1958, Meddelanden fran Lunds Astronomiska Observatorium Serie
  II, 136, 1

\bibitem[{{Huang} {et~al.}(2012){Huang}, {Haynes}, {Giovanelli}, \&
  {Brinchmann}}]{2012huang}
{Huang}, S., {Haynes}, M.~P., {Giovanelli}, R., \& {Brinchmann}, J. 2012, \apj,
  756, 113

\bibitem[{{Impey} {et~al.}(1988){Impey}, {Bothun}, \& {Malin}}]{1988impey}
{Impey}, C., {Bothun}, G., \& {Malin}, D. 1988, \apj, 330, 634

\bibitem[{{Infante-Sainz} {et~al.}(2019){Infante-Sainz}, {Trujillo}, \&
  {Rom{\'a}n}}]{2019raul}
{Infante-Sainz}, R., {Trujillo}, I., \& {Rom{\'a}n}, J. 2019, arXiv e-prints,
  arXiv:1911.01430

\bibitem[{{Koda} {et~al.}(2015){Koda}, {Yagi}, {Yamanoi}, \&
  {Komiyama}}]{2015koda}
{Koda}, J., {Yagi}, M., {Yamanoi}, H., \& {Komiyama}, Y. 2015, \apjl, 807, L2

\bibitem[{{Kravtsov}(2013)}]{2013kravtsov}
{Kravtsov}, A.~V. 2013, \apjl, 764, L31

\bibitem[{{Leroy} {et~al.}(2008){Leroy}, {Walter}, {Brinks}, {Bigiel}, {de
  Blok}, {Madore}, \& {Thornley}}]{2008leroy}
{Leroy}, A.~K., {Walter}, F., {Brinks}, E., {et~al.} 2008, AJ, 136, 2782

\bibitem[{{Mancera Pi{\~n}a} {et~al.}(2018){Mancera Pi{\~n}a}, {Peletier},
  {Aguerri}, {Venhola}, {Trager}, \& {Choque Challapa}}]{2018pina}
{Mancera Pi{\~n}a}, P.~E., {Peletier}, R.~F., {Aguerri}, J.~A.~L., {et~al.}
  2018, \mnras, 481, 4381

\bibitem[{{Mart{\'\i}nez-Lombilla} {et~al.}(2019){Mart{\'\i}nez-Lombilla},
  {Trujillo}, \& {Knapen}}]{2019lombilla}
{Mart{\'\i}nez-Lombilla}, C., {Trujillo}, I., \& {Knapen}, J.~H. 2019, MNRAS,
  483, 664

\bibitem[{{Mihos} {et~al.}(2015){Mihos}, {Durrell}, {Ferrarese}, {Feldmeier},
  {C{\^o}t{\'e}}, {Peng}, {Harding}, {Liu}, {Gwyn}, \&
  {Cuillandre}}]{2015mihos}
{Mihos}, J.~C., {Durrell}, P.~R., {Ferrarese}, L., {et~al.} 2015, \apjl, 809,
  L21

\bibitem[{{Miller} {et~al.}(2019){Miller}, {van Dokkum}, {Mowla}, \& {van der
  Wel}}]{2019miller}
{Miller}, T.~B., {van Dokkum}, P., {Mowla}, L., \& {van der Wel}, A. 2019,
  \apjl, 872, L14

\bibitem[{{Mu{\~n}oz} {et~al.}(2015){Mu{\~n}oz}, {Eigenthaler}, {Puzia},
  {Taylor}, {Ordenes-Brice{\~n}o}, {Alamo-Mart{\'\i}nez}, {Ribbeck},
  {{\'A}ngel}, {Capaccioli}, {C{\^o}t{\'e}}, {Ferrarese}, {Galaz}, {Hempel},
  {Hilker}, {Jord{\'a}n}, {Lan{\c{c}}on}, {Mieske}, {Paolillo}, {Richtler},
  {S{\'a}nchez-Janssen}, \& {Zhang}}]{2015munoz}
{Mu{\~n}oz}, R.~P., {Eigenthaler}, P., {Puzia}, T.~H., {et~al.} 2015, \apjl,
  813, L15

\bibitem[{{Nair} {et~al.}(2011){Nair}, {van den Bergh}, \&
  {Abraham}}]{2011nair}
{Nair}, P., {van den Bergh}, S., \& {Abraham}, R.~G. 2011, \apjl, 734, L31

\bibitem[{{Nair} \& {Abraham}(2010)}]{2010preethi}
{Nair}, P.~B. \& {Abraham}, R.~G. 2010, \apjs, 186, 427

\bibitem[{{Prole} {et~al.}(2019){Prole}, {van der Burg}, {Hilker}, \&
  {Davies}}]{2019prole}
{Prole}, D.~J., {van der Burg}, R.~F.~J., {Hilker}, M., \& {Davies}, J.~I.
  2019, \mnras, 488, 2143

\bibitem[{{Redman}(1936)}]{1936redman}
{Redman}, R.~O. 1936, \mnras, 96, 588

\bibitem[{{Roediger} \& {Courteau}(2015)}]{2015roediger}
{Roediger}, J.~C. \& {Courteau}, S. 2015, \mnras, 452, 3209

\bibitem[{{Rom{\'a}n} \& {Trujillo}(2017{\natexlab{a}})}]{2017roman}
{Rom{\'a}n}, J. \& {Trujillo}, I. 2017{\natexlab{a}}, \mnras, 468, 703

\bibitem[{{Rom{\'a}n} \& {Trujillo}(2017{\natexlab{b}})}]{2017s82udgs}
{Rom{\'a}n}, J. \& {Trujillo}, I. 2017{\natexlab{b}}, \mnras, 468, 4039

\bibitem[{{Rom{\'a}n} \& {Trujillo}(2018)}]{2018s82rectified}
{Rom{\'a}n}, J. \& {Trujillo}, I. 2018, Research Notes of the American
  Astronomical Society, 2, 144

\bibitem[{{Ruiz-Lara} {et~al.}(2018){Ruiz-Lara}, {Beasley},
  {Falc{\'o}n-Barroso}, {Rom{\'a}n}, {Pinna}, {Brook}, {Di Cintio},
  {Mart{\'\i}n-Navarro}, {Trujillo}, \& {Vazdekis}}]{2018tomas}
{Ruiz-Lara}, T., {Beasley}, M.~A., {Falc{\'o}n-Barroso}, J., {et~al.} 2018,
  \mnras, 478, 2034

\bibitem[{{Sandage} \& {Binggeli}(1984)}]{1984bruno}
{Sandage}, A. \& {Binggeli}, B. 1984, \aj, 89, 919

\bibitem[{{Shen} {et~al.}(2003){Shen}, {Mo}, {White}, {Blanton}, {Kauffmann},
  {Voges}, {Brinkmann}, \& {Csabai}}]{2003shen}
{Shen}, S., {Mo}, H.~J., {White}, S.~D.~M., {et~al.} 2003, \mnras, 343, 978

\bibitem[{{Strauss} {et~al.}(2002){Strauss}, {Weinberg}, {Lupton}, {Narayanan},
  {Annis}, {Bernardi}, {Blanton}, {Burles}, {Connolly}, {Dalcanton}, {Doi},
  {Eisenstein}, {Frieman}, {Fukugita}, {Gunn}, {Ivezi{\'c}}, {Kent}, {Kim},
  {Knapp}, {Kron}, {Munn}, {Newberg}, {Nichol}, {Okamura}, {Quinn}, {Richmond},
  {Schlegel}, {Shimasaku}, {SubbaRao}, {Szalay}, {Vanden Berk}, {Vogeley},
  {Yanny}, {Yasuda}, {York}, \& {Zehavi}}]{2002strauss}
{Strauss}, M.~A., {Weinberg}, D.~H., {Lupton}, R.~H., {et~al.} 2002, \aj, 124,
  1810

\bibitem[{Teeninga {et~al.}(2016)Teeninga, Moschini, C.~Trager, \&
  Wilkinson}]{2016mto}
Teeninga, P., Moschini, U., C.~Trager, S., \& Wilkinson, M. 2016, 1

\bibitem[{{Trujillo} {et~al.}(2019){Trujillo}, {Beasley}, {Borlaff},
  {Carrasco}, {Di Cintio}, {Filho}, {Monelli}, {Montes}, {Rom{\'a}n},
  {Ruiz-Lara}, {S{\'a}nchez Almeida}, {Valls-Gabaud}, \&
  {Vazdekis}}]{2019trujillo}
{Trujillo}, I., {Beasley}, M.~A., {Borlaff}, A., {et~al.} 2019, \mnras, 486,
  1192

\bibitem[{{Trujillo} {et~al.}(2017){Trujillo}, {Roman}, {Filho}, \&
  {S{\'a}nchez Almeida}}]{2017ugc2162}
{Trujillo}, I., {Roman}, J., {Filho}, M., \& {S{\'a}nchez Almeida}, J. 2017,
  \apj, 836, 191

\bibitem[{{van der Burg} {et~al.}(2016){van der Burg}, {Muzzin}, \&
  {Hoekstra}}]{2016vanderburg}
{van der Burg}, R. F.~J., {Muzzin}, A., \& {Hoekstra}, H. 2016, \aap, 590, A20

\bibitem[{{van Dokkum} {et~al.}(2018){van Dokkum}, {Danieli}, {Cohen},
  {Merritt}, {Romanowsky}, {Abraham}, {Brodie}, {Conroy}, {Lokhorst}, {Mowla},
  {O'Sullivan}, \& {Zhang}}]{2018naturedok}
{van Dokkum}, P., {Danieli}, S., {Cohen}, Y., {et~al.} 2018, \nat, 555, 629

\bibitem[{{van Dokkum} {et~al.}(2015{\natexlab{a}}){van Dokkum}, {Abraham},
  {Merritt}, {Zhang}, {Geha}, \& {Conroy}}]{2015giantgalaxies}
{van Dokkum}, P.~G., {Abraham}, R., {Merritt}, A., {et~al.} 2015{\natexlab{a}},
  \apjl, 798, L45

\bibitem[{{van Dokkum} {et~al.}(2015{\natexlab{b}}){van Dokkum}, {Romanowsky},
  {Abraham}, {Brodie}, {Conroy}, {Geha}, {Merritt}, {Villaume}, \&
  {Zhang}}]{2015df44}
{van Dokkum}, P.~G., {Romanowsky}, A.~J., {Abraham}, R., {et~al.}
  2015{\natexlab{b}}, \apjl, 804, L26

\bibitem[{{Venhola} {et~al.}(2017){Venhola}, {Peletier}, {Laurikainen}, {Salo},
  {Lisker}, {Iodice}, {Capaccioli}, {Verdois Kleijn}, {Valentijn}, {Mieske},
  {Hilker}, {Wittmann}, {van de Ven}, {Grado}, {Spavone}, {Cantiello},
  {Napolitano}, {Paolillo}, \& {Falc{\'o}n-Barroso}}]{2017venhola}
{Venhola}, A., {Peletier}, R., {Laurikainen}, E., {et~al.} 2017, \aap, 608,
  A142

\end{thebibliography}

\appendix 

\onecolumn 
\section{Table of measurements}

\label{sec:table}
\begin{table}[h]
\centering
\caption{Measured parameters for the sample of UDGs. We include the common name, position, axis ratio (q), and the position angle (PA) of the ellipses used to extract the surface brightness profiles (measured counter clockwise starting from the horizontal axis), the spectroscopic redshifts $z$ (if available) of the galaxies \citep{2017s82udgs, 2017ugc2162, 2015df44, 2019emsellem}, Galactic extinctions both in the g and r bands (from NED), effective radius $R_{\rm e}$ (measured in the $g$-band), the radial location $R_1$ of the isomass contour at 1\,$M_{\odot}/$pc$^2$, the Holmberg Radius ($R_{\rm H}$; defined here using the isophote at 26\,mag/arcsec$^2$ in the $g$-band) and the stellar mass of galaxies (assuming a Chabrier IMF). The quantities are given showing only the significant figures up to which the values can be regarded reliable. The table for the dwarf galaxies is available in the online version of TCK19.}
\label{table:parameters} 
\begin{tabular}{cccccccccccc}

\hline \\
\textbf{Name} & \textbf{R.A.} & \textbf{Dec}&\textbf{q} &\textbf{PA} &  \textbf{z}& \textbf{A$_g$}  & \textbf{A$_r$}  & \textbf{$R_{\rm e}$} & \textbf{$R_1$} & \textbf{$R_{\rm H}$} & \textbf{Log($M_\star$/$M_{\odot}$)}  \\  
 & \textbf{(deg)} &\textbf{(deg)}& &\textbf{(deg)} & &\textbf{(mag)}  & \textbf{(mag)} &  \textbf{(kpc)} & \textbf{(kpc)} & \textbf{(kpc)} & \\  

\hline \\

DF44 &  195.24167& 26.97638& 0.69&      60.0 &0.0231&   0.034&  0.024   &3.31&  4.34&   3.33&   8.16\\
{[KKS2000]}04 & 40.44500&  $-$8.40258&     0.87&   40.0 & 0.0060   &0.081& 0.056&  1.30&   2.62&   1.69& 7.86\\
UGC2162 &40.09625&      1.22917&        0.77&   25.0 &  0.0039& 0.117&  0.081&  2.23&   2.22&   2.91&   7.58\\
UDG-B1 &        50.08800&       $-$1.17000&     0.46&   95.0 &  0.0212& 0.213&  0.148&  3.93&   4.11&   6.94&   7.90\\
UDG-B3 &        49.96000&       $-$0.85500&     0.86&   41.0 &  0.0212& 0.195&  0.135&  3.26&   4.91&   5.17&   8.41\\
UDG-B2 &        9.60000&        1.10600&        0.50& 7.0&      0.0141& 0.060&  0.041&  1.76&   1.92&   3.28&   7.29\\
UDG-B4 &        9.88900&        1.11500&        0.67&   107.0&  0.0141& 0.064&  0.044&  1.49&   2.25&   2.57&   7.51\\
UDG-B5 &        9.96900&        0.38300&        0.75&   15.0 &  0.0141& 0.058&  0.040&  2.95&   4.06&   4.64&   8.11\\
UDG-R1 &        9.97400&        0.80800&        0.64&   167.0&  0.0141& 0.065&  0.045&  1.26&   1.55&   0.61&   7.07\\
UDG-R4 &        9.76900&        1.09900&        0.63&   10.0&   0.0141& 0.059&  0.041&  1.52&   1.61&   1.22&   7.30\\
UDG-R5 &        358.61600&      0.32200&        0.59&   120.0&  0.0266& 0.135&  0.094&  2.07&   2.65&   1.96&   7.51\\
UDG-R6 &        358.49700&      0.45400&        0.73&   30.0&   0.0266& 0.135&  0.093&  1.78&   3.95&   2.55&   8.03\\

\hline \\
\end{tabular}
\end{table}
\twocolumn
\section{Comparison with an isophotal size indicator: The Holmberg radius}

\label{sec:holmbergradius}

\begin{figure}
    \centering
        \includegraphics[width=1.0\columnwidth]{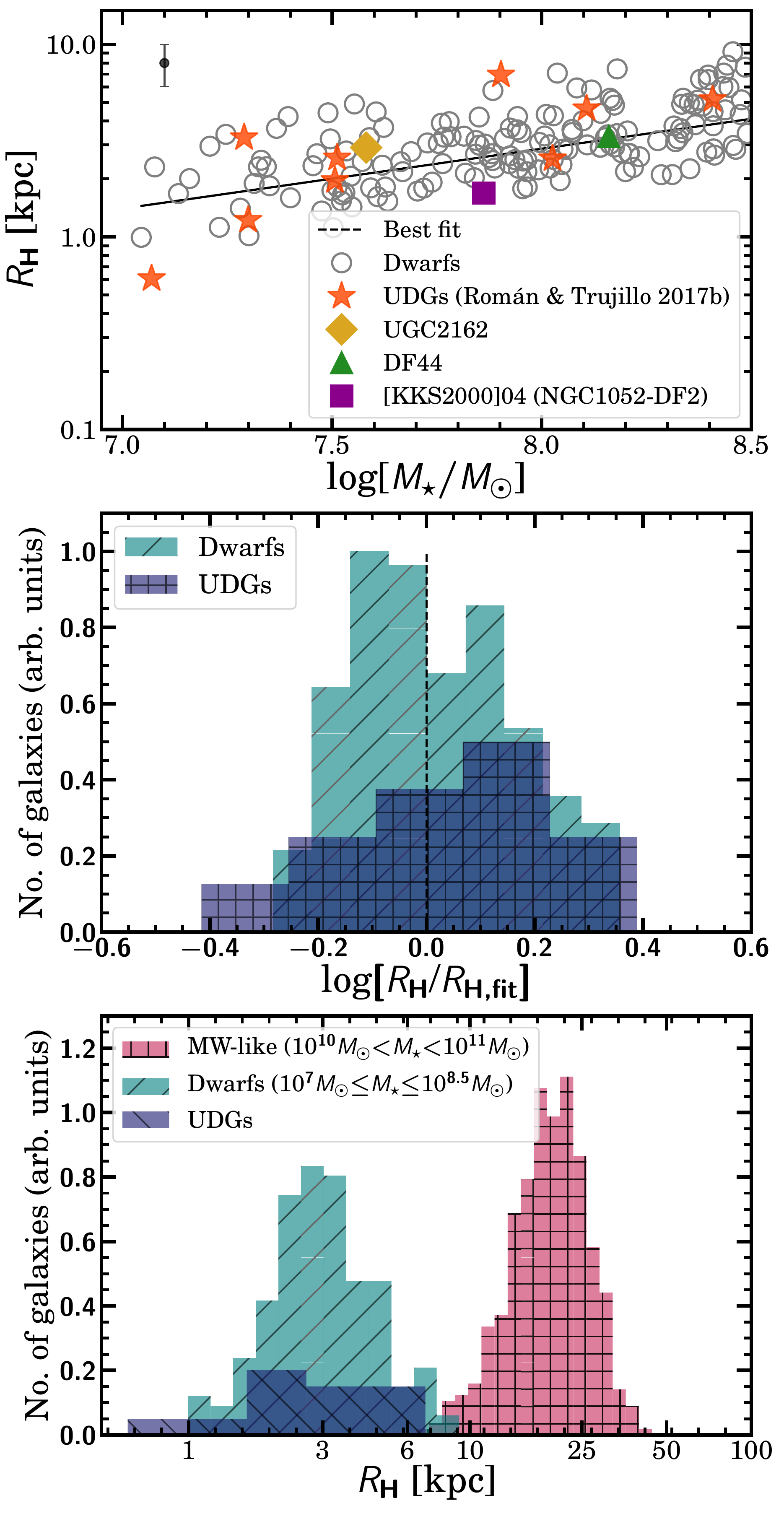}
    \caption{Distribution of UDGs and dwarf galaxies using the Holmberg radius. \textit{Top}: Holmberg radius ($R_{\rm H}$)--stellar mass plane for dwarfs (grey) and UDGs (colours). The best fit line of the relation for the dwarf sample is also over-plotted. The upper left corner of the plot shows the typical uncertainty in our measurements (see TCK19). \textit{Middle}: Histogram showing the distribution of  $R_{H}/R_{H,\rm fit}$ where `fit' refers to the best-fit line of each relation for the dwarf sample. \textit{Bottom}: Histogram showing the $R_{\rm H}$ distribution of UDGs, dwarfs, and MW-like galaxies.}
    \label{fig:rholmberg}
\end{figure}

Similar to Figs. \ref{fig:relation} and \ref{fig:histograms}, Fig. \ref{fig:rholmberg} shows the distribution of UDGs, dwarfs, and MW-like galaxies using the Holmberg radius. Here we define the Holmberg radius using the isophote at 26 mag/arcsec$^2$ in the $g$-band. The middle panel clearly demonstrates that UDGs have sizes that are within the size range of dwarf galaxies. The KS test using $R_{H}/R_{H,\rm fit}$ gives a p-value of 0.54. Finally, the lower panel shows that the sizes of UDGs are not compatible with those of MW-like galaxies. Therefore, our conclusions are further reinforced by taking a widely used isophotal size indicator. In other words, our conclusions regarding the sizes of UDGs are not related with our specific size definition to describe the extensions of galaxies.

\section{Other stellar mass density proxies to measure the size of dwarfs and UDGs}

\label{sec:othersize}

\begin{figure}[h]
    \centering
        \includegraphics[width=1.0\columnwidth]{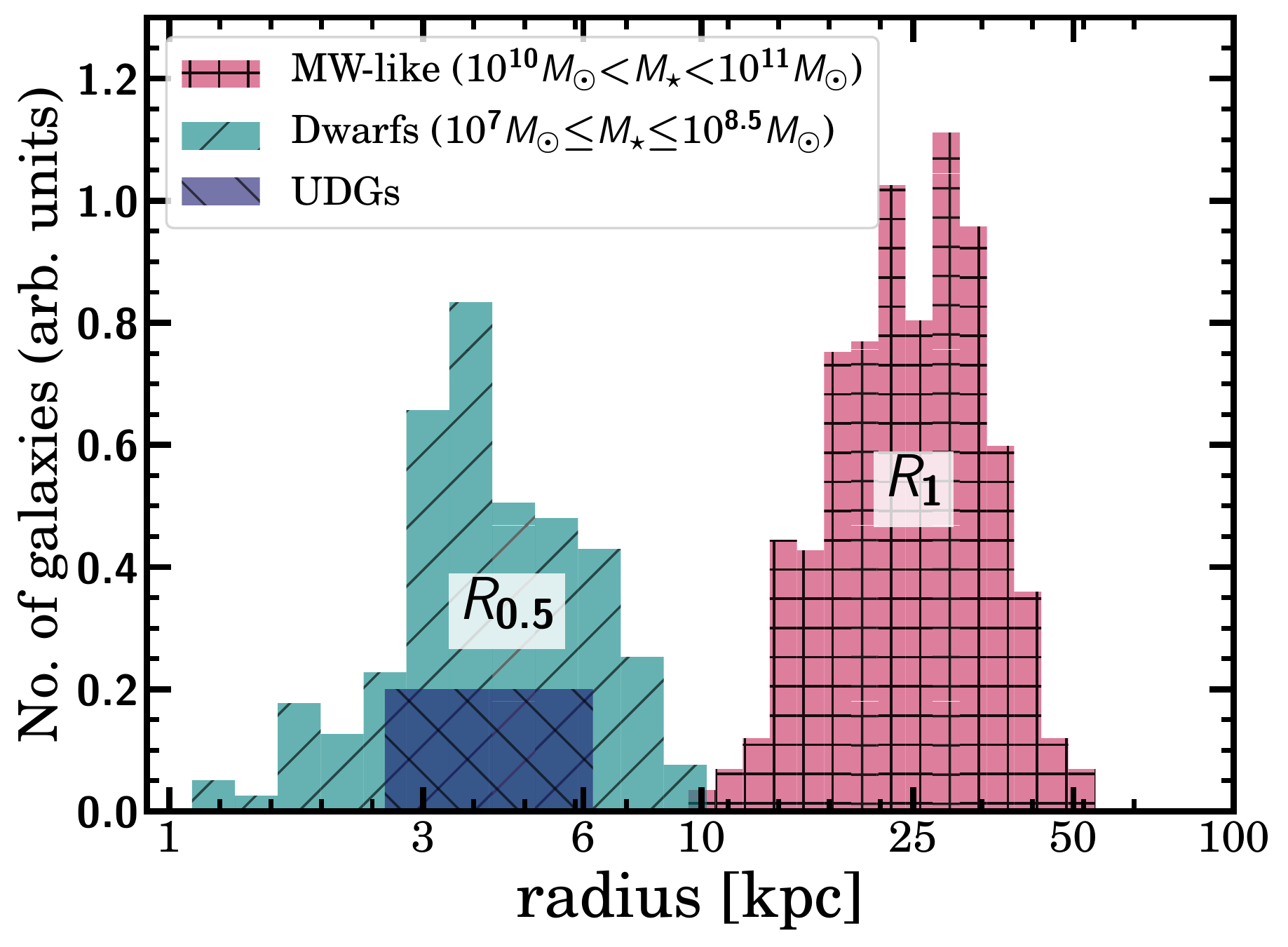}
    \caption{Histograms showing the size distribution of UDGs ($R_{0.5}$), dwarfs ($R_{0.5}$) and MW-like galaxies ($R_1$).}
    \label{fig:rhalf}
\end{figure}

\normalsize{In TCK19, we proposed a size indicator based on the location of the gas density threshold for star formation in galaxies. On both theoretical and observational grounds, we selected the location of an isomass contour at 1 $M_{\odot}/$pc$^2$ as a proxy for such a value. While this value is motivated by the location of the disc truncation in MW-like galaxies \citep[see][]{2019lombilla}, for galaxies such as dwarfs where the level of star formation is lower, a better proxy for their gas density threshold could be given by a lower isomass contour \citep{2008leroy, 2012huang}. Therefore, we repeated our analysis using an isomass contour at 0.5\,$M_{\odot}$/pc$^2$ instead of 1\,$M_{\odot}$/pc$^2$ as a galaxy size indicator for UDGs and dwarfs (see Fig. \ref{fig:egprofiles} for two examples). We call this size parameter $R_{0.5}$. \par 
Figure \ref{fig:rhalf} shows the size distribution of UDGs and dwarfs using $R_{0.5}$ and that of MW-like systems using $R_1$. Although $R_{0.5}$ increases the sizes of the UDGs and dwarf galaxies as expected, their extensions never reach those of MW-like galaxies. Similar to the results shown in Fig. \ref{fig:histograms} (right panel), the null hypothesis is completely rejected upon comparing the size distributions of UDGs and MW-like galaxies. Therefore, our conclusion that UDGs do not have sizes comparable to MW-like galaxies} remains unchanged even with the use of another gas density threshold for size.

\end{document}